\documentclass[a4paper,11pt]{article}

\usepackage{jcappub}
\usepackage{slashed}
\usepackage{bm}
\usepackage{latexsym,amssymb,amsmath,float,url,mathrsfs}
\usepackage{latexsym}
\usepackage{graphicx}
\usepackage{epstopdf}
\usepackage{amsfonts}
\usepackage{amsmath}
\usepackage{amssymb}
\usepackage{comment}
\usepackage{subfigure}
\usepackage{natbib}
\usepackage{hyperref}
\usepackage{pifont}
\usepackage{blindtext}
\usepackage{adjustbox}
\usepackage{multirow}
\usepackage{tabularx}
\usepackage[table]{xcolor}
\usepackage{orcidlink}
\usepackage{tikz}
\usetikzlibrary{tikzmark}
\usepackage{amssymb}
\usepackage{bbold}

\usepackage{blkarray}
\usepackage{graphicx}
\usepackage{amsfonts}
\usepackage{soul}
\usepackage{amssymb}
\usepackage{amsmath}
\usepackage{cancel}
\usepackage{tcolorbox}

\usepackage{graphics,appendix,afterpage,makecell} 

\def\abs#1{\left| #1\right|}

\definecolor{oucrimsonred}{rgb}{0.6, 0.0, 0.0}
\definecolor{persianblue}{rgb}{0.11, 0.22, 0.73}
\definecolor{forestgreen}{rgb}{0.13,0.35,0.13}
\definecolor{lightgray}{rgb}{0.83, 0.83, 0.83}
 \hypersetup{colorlinks, citecolor=oucrimsonred, linkcolor=black, urlcolor=oucrimsonred}
\definecolor{cornellred}{rgb}{0.7, 0.11, 0.11}
\definecolor{navyblue}{rgb}{0.0, 0.0, 0.5}
\definecolor{amethyst}{rgb}{0.6, 0.4, 0.8}
\definecolor{yellow}{rgb}{1.0, 1.0, 0.0}
\definecolor{firebrick}{rgb}{0.7, 0.13, 0.13}
\definecolor{tangerineyellow}{rgb}{1.0, 0.8, 0.0}
\definecolor{deepfuchsia}{rgb}{0.76, 0.33, 0.76}
\definecolor{amber}{rgb}{1.0, 0.75, 0.0}
\definecolor{VioletRed4}{rgb}{0.55, 0.13, .32}
\definecolor{indiagreen}{rgb}{0.07, 0.53, 0.03}
\definecolor{VioletRed4}{rgb}{0.55, 0.13, .32}
\newcommand{\be}{\begin{equation}}
\newcommand{\ee}{\end{equation}}
\newcommand{\bea}{\begin{equation} \begin{aligned}}
\newcommand{\eea}{\end{aligned} \end{equation}}

\definecolor{oucrimsonred}{rgb}{0.6, 0.0, 0.0}
\newcommand\vertarrowbox[3][6ex]{%
  \begin{array}[t]{@{}c@{}} #2 \\
  \left\uparrow\vcenter{\hrule height #1}\right.\kern-\nulldelimiterspace\\
  \makebox[0pt]{\scriptsize#3}
  \end{array}%
}

\definecolor{verdechiaro}{rgb}{0.6,1,0.6}
\definecolor{giallochiaro}{rgb}{1,1,0.6}
\definecolor{bluscuro}{rgb}{0.15, 0.2, 0.9}
\definecolor{verdes}{rgb}{0.1, 0.5, 0.1}%
\definecolor{tangerineyellow}{rgb}{1.0, 0.8, 0.0}

\definecolor{americanrose}{rgb}{1.0, 0.01, 0.24}
\definecolor{cobalt}{rgb}{0.0, 0.28, 0.67}
\definecolor{brandeisblue}{rgb}{0.0, 0.44, 1.0}
\definecolor{mycolor}{rgb}{0.0, 0.0, 0.5}
\definecolor{oxfordblue}{rgb}{0.0, 0.13, 0.28}
\definecolor{azure}{rgb}{0.0, 0.5, 1.0}
\definecolor{turquoiseblue}{rgb}{0.0, 1.0, 0.94}
\newtcolorbox{mynewbox}[1]{colback=white!5!white,colframe=azure!75!black,fonttitle=\bfseries,title=#1}
\newtcolorbox{mybox}{colback=mycolor!5!white,colframe=azure!75!black}
\newtcolorbox{mynamedbox}[1]{colback=mycolor!5!white,colframe=azure!75!black,title=#1}
\definecolor{venetianred}{rgb}{0.78, 0.03, 0.08}
\newtcolorbox{mynamedbox1}[1]{colback=venetianred!5!white,colframe=venetianred!80!black,title=#1}
\newtcolorbox{mynamedbox2}[1]{colback=azure!5!white,colframe=azure!80!black,title=#1}

\newcommand{\td}{{\rm d}}

\definecolor{verdes}{rgb}{0.1, 0.5, 0.1}%
\definecolor{cornellred}{rgb}{0.7, 0.11, 0.11}

\definecolor{VioletRed4}{rgb}{0.55, 0.13, .32}

\hypersetup{
     colorlinks   = true,
     citecolor    = violet,
     urlcolor     = violet,
     linkcolor    = violet}

\definecolor{rossocorsa}{rgb}{0.83, 0.0, 0.0}

\usepackage[normalem]{ulem}

\title{
The Vanishing of the  Non-linear Static  Love Number  of  Kerr Black Holes and the Role of Symmetries}

\author[a]{L.-R. Gounis}
\author[a,b]{A. Kehagias}
\author[b,c]{A. Riotto}
\affiliation[a]{Physics Division, National Technical University of Athens, Athens, 15780, Greece}
\affiliation[b]{Department of Theoretical Physics, 
24 quai E. Ansermet, CH-1211 Geneva 4, Switzerland}

\affiliation[c]{ Gravitational Wave Science Center,  
24 quai E. Ansermet, CH-1211 Geneva 4, Switzerland}

\abstract{
We investigate the tidal response of Kerr black holes in four-dimensional spacetimes subjected to external gravitational fields. Using the Ernst formalism and Weyl coordinates, we analyze the non-linear tidal deformation of rotating black holes and demonstrate that their static tidal Love numbers vanish at all orders of the external tidal field.  We also show that this result is intimately related to the presence of   underlying non-linear symmetries. Our analysis generalizes previous findings for Schwarzschild black holes and confirms the robustness of four-dimensional black holes against tidal forces.

}

\emailAdd{el190810@central.ntua.gr}
\emailAdd{kehagias@central.ntua.gr}
\emailAdd{antonio.riotto@unige.ch}

\begin{document}
\maketitle

\section{Introduction}
Gravitational Waves (GWs) and Black Holes (BHs) are key predictions of General Relativity (GR), validated by groundbreaking observations such as the detection of GWs from BH mergers by the LIGO and Virgo collaborations~\cite{LIGOScientific:2021sio}. These detections have provided critical evidence supporting Einstein theory of gravity, showing no evidence for deviations from it \cite{Kehagias:2024yyp}.

During the inspiral phase of a compact binary system, such as those involving neutron stars or BHs, tidal interactions become significant when the orbital separation is sufficiently small. These tidal effects influence both the system's dynamics and the emitted GWs. The interplay between GWs and tidal effects is essential for refining binary inspiral models and testing GR under extreme conditions.

Tidal effects are characterized by parameters known as  Love Numbers, which quantify an object's deformation in response to the gravitational field of its companion. In particular the static Tidal Love Numbers (TLNs) depend on the internal structure and composition of the compact objects undergoing tidal deformation~\cite{poisson_will_2014}. These parameters play a pivotal role in modifying the gravitational waveform, with their contributions emerging at the fifth post-Newtonian order~\cite{Flanagan:2007ix}. For example, the nonzero TLNs of neutron stars provide valuable insights into the equation of state of dense nuclear matter. In contrast, BHs are expected to have zero TLNs due to their lack of a rigid structure. This result is typically demonstrated using perturbation theory, showing that a linear tidal deformation with amplitude proportional to \( r^\ell \) does not elicit an \( r^{-\ell-1} \) response ($\ell$ being the corresponding multipole), resulting in vanishing static TLNs. Linear perturbations induced by external tidal forces cannot produce nonzero TLNs~\cite{Binnington:2009bb,Damour:2009vw,Damour:2009va,Pani:2015hfa,Pani:2015nua,Porto:2016zng,LeTiec:2020spy,Chia:2020yla,LeTiec:2020bos,Poisson:2021yau,Kehagias:2024yzn}. This phenomenon appears to stem from underlying hidden symmetries~\cite{Hui:2020xxx,Charalambous:2021mea,Charalambous:2021kcz,Hui:2021vcv,Hui:2022vbh,Charalambous:2022rre,Ivanov:2022qqt,Katagiri:2022vyz,Bonelli:2021uvf,Kehagias:2022ndy,BenAchour:2022uqo,Berens:2022ebl,DeLuca:2023mio,Rai:2024lho}.

Recent analyses have confirmed that static TLNs also vanish for second-order perturbations in the external tidal field~\cite{Riva:2023rcm,Riva:2024}. Furthermore, for the Schwarzschild BH, the vanishing of TLNs has been proven to hold for the parity-even perturbations at all orders in the external tidal field~\cite{Kehagias:2024rtz,Combaluzier-Szteinsznaider:2024sgb}.

The fact  that the static TLN for BHs vanishes or not is of primary importance to distinguish BH mergers  from neutron star mergers \cite{Crescimbeni:2024cwh}, having neutron stars a sizeable TLN. Furthermore, even the the merger of two spinless BHs give rise to a spinning Kerr BH.  This calls for a unavoidable question: do Kerr BHs have a vanishing static TLN at any order in the external tidal force?

The case of rotating BHs, modeled by the Kerr solution, presents additional challenges. Rotation introduces frame-dragging effects and modifies the geometry of the spacetime, complicating the analysis of tidal interactions. Understanding the tidal response of Kerr BHs is essential, not only for theoretical completeness, but also for modeling gravitational waveforms from realistic astrophysical systems, where BHs are often expected to spin. 

In this paper, we address this question of the vanishing of the static TLN of kerr BHs by employing the Ernst formalism ~\cite{Ernst:1967wx} and Weyl coordinates to analyze the tidal response of Kerr BHs. The Ernst potential provides a powerful framework for describing axially symmetric spacetimes, allowing us to incorporate rotation and non-linear effects systematically. By expressing the Kerr metric in prolate spheroidal coordinates, we generalize previous results for Schwarzschild BHs and demonstrate that the static tidal Love numbers of Kerr BHs vanish at all orders in the external tidal field. We will also identify   the non-linear symmetries responsible for such a result.

This result highlights the robustness of the symmetry-based arguments that govern BH responses and underscores the distinctive nature of BHs as solutions to GR. The vanishing TLNs reaffirm the principle that BHs, unlike other compact objects, do not retain any permanent deformation under static tidal forces. This study contributes to the broader understanding of BH physics, offering new perspectives on their interaction with external fields and implications for gravitational wave astronomy.

The paper is organized as follows: Section 2 reviews the Weyl class of static, axisymmetric vacuum solutions and introduces the Ernst potential formalism. Section 3 revisits the tidal response of Schwarzschild BHs, establishing the framework for non-linear tidal effects. Section 4 extends the analysis to Kerr BHs, detailing the transition to prolate spheroidal coordinates and examining the decaying and growing quadrupole modes. Section 5 investigates the impact of non-linear tidal interactions and their role in ensuring the vanishing of TLNs. Section 6 discusses the role played by the non-linear symmetries. Section 7 concludes with a discussion of the implications and potential extensions of this work. Finally, Appendices  A and B discuss the transition to Boyer-Lindquist coordinates and other multipole basis, offering a complementary
perspective.

\section{The Weyl class of static, axisymmetric vacuum solutions}
As demonstrated by Ernst ~\cite{Ernst:1967wx}, the field equations for a uniformly rotating, axially symmetric source can be reformulated using a simple variational principle. Following this approach unified solutions for Weyl and Papapetrou metrics emerge providing us with a direct derivation of the Schwarzschild as well as the Kerr metric in prolate spheroidal coordinates. New solutions for the case of Kerr BH in tidal environments can also be obtained in this way, allowing us to make statements about the non-linear static love numbers of Kerr BHs.
We can start our analysis by considering a static axisymmetric Weyl metric in the following form~\cite{Papapetrou:1953zz}
\begin{equation}
    {\rm d}s^2 = f^{-1} \left[ e^{2\gamma} ({\rm d}\rho^2 + {\rm d}z^2) + \rho^2 {\rm d}\varphi^2 \right] - f ({\rm d}t -\omega {\rm d}\varphi)^2, \label{metric}
\end{equation}
where $f=f(\rho,z)$, $\omega=\omega(\rho,z)$ and $\gamma=\gamma(\rho,z)$. It turns out that that the equations for  $f$ and $\omega$  which follow from the  vacuum Einstein field equations  ($R_{\mu\nu}$=0) can be decoupled from the equation for $\gamma(\rho,z)$ and are given by
\begin{equation}
    f \nabla^2 f = \nabla f \cdot \nabla f - \rho^{-2} f^4 \nabla \omega \cdot \nabla \omega, \label{eq1}
\end{equation}
\begin{equation}
    \nabla \cdot \left(\rho^{-2} f^2 \nabla \omega\right) = 0. \label{eq2}
\end{equation}
We may now introduce a new scalar $\phi$ from $\omega$ as 
\begin{equation}
    \nabla \phi=-\frac{f^2}{\rho}\hat{n}_\varphi \times \nabla \omega. \label{phi}
\end{equation}
Moreover, the equations for the third function $\gamma(r,\theta)$ are written in terms of the $\mathcal{E}$ as~\cite{Ernst:2006yg}
\begin{align}
    \gamma_{,z} &= \frac{1}{4} \rho f^{-2} \left[(\mathcal{E}_{,\rho} )(\mathcal{E}^{*}_{,z} ) 
+ (\mathcal{E}_{,z} )(\mathcal{E}^{*}_{,\rho} )\right] ,\notag \\
\gamma_{,\rho} &= \frac{1}{4} \rho f^{-2} \left[(\mathcal{E}_{,\rho})(\mathcal{E}^{*}_{,\rho}) 
- (\mathcal{E}_{,z})(\mathcal{E}^{*}_{,z} )\right]  .\label{gamma}
\end{align}
We now introduce prolate spheroidal coordinates ($t, x, y, \varphi$) instead of Weyl
coordinates by writing~\cite{Zipoy:1966btu,Quevedo:1989rfm}
\begin{align}
    \rho &= \rho_0 (x^2 - 1)^{1/2} (1 - y^2)^{1/2}, \quad x \geq 1, \quad |y| \leq 1 ,\nonumber\\
    z &= \rho_0 x y, \quad \rho_0 = \text{constant}. \label{pro}
\end{align}
We will see later that $\rho_0$ is related to the mass and the spin parameter of the BHs we are interested in describing.
In such coordinates, the metric in
Eq. (2.1) is written as
\begin{equation}
\label{metricx}
    {\rm d}s^2 =  \rho_0^2 f^{-1} \left[ e^{2\gamma}(x^2 - y^2) \left( \frac{{\rm d}x^2}{x^2 - 1} + \frac{{\rm d}y^2}{1 - y^2} \right) + (x^2 - 1)(1 - y^2) {\rm d}\phi^2 \right] - f ({\rm d}t - \omega {\rm d}\phi)^2.
\end{equation}
Furthermore, for later use,  the differential operators we previously introduced  are written in prolate spheroidal coordinates  now take the following form:
\begin{align}
    \nabla &\equiv \rho_0^{-1} (x^2 - y^2)^{-1/2} \left[ \hat{n}_x (x^2 - 1)^{1/2} \partial_x + \hat{n}_y (1 - y^2)^{1/2} \partial_y \right], \nonumber\\
    \nabla^2 &= \rho_0^{-2} (x^2 - y^2)^{-1} \bigg\{ \partial_x \Big[ (x^2 - 1) \partial_x \Big] + \partial_y \Big[ (1 - y^2) \partial_y \Big] \bigg\} ,
    \end{align}
    whereas, the inner product of the gradients of two functions $A$ and $B$ is 
    \begin{align}
    \nabla A\cdot\nabla B&= \rho_0^{-2} (x^2 - y^2)^{-1} \Big[  (x^2 - 1) \partial_xA\,\partial_xB+  (1 - y^2) \partial_yA\,\partial_yB \Big].\nonumber
\end{align}
It has been shown ~\cite{Ernst:1967wx} that Eqs. (\ref{eq1}) and (\ref{eq2})  can also be obtained through a complex function, the Ernst potential $\mathcal{E}$, defined as 
\begin{equation}
    \mathcal{E}=f+i\phi.
\end{equation}
In particular, Eqs. (\ref{eq1}) and (\ref{eq2}), are equivalent to the equation of motion for the Ernst potential $\mathcal{E}$ which are derived from the action 
\begin{equation}
    \mathcal{S}_{\mathcal{E}}=\int \frac{\nabla \mathcal{E}\cdot \nabla\mathcal{E}^*}{(\mathcal{E}+\mathcal{E}^*)^2} d^2 x, \label{ee}
\end{equation}
so that  the corresponding equations 
\begin{equation}
\left(\mathcal{E}+\mathcal{E}^*\right)\nabla^2\mathcal{E}-\nabla \mathcal{E}\cdot \nabla \mathcal{E}=0, \label{eqe}
\end{equation}
reproduce Eqs. (\ref{eq1}) and (\ref{eq2}). As a result, the problem of finding axisymmetric, stationary vacuum solutions to the Einstein equations is in fact reduced to appropriately solve Eq. (\ref{eqe}) for the Ernst potential $\mathcal{E}$.

\section{The Schwarzschild BH in external tidal fields}
Although the Schwarzschild BH in external tidal fields has been extensively discussed  in Ref.
~\cite{Kehagias:2024rtz}, let us recall here its description in terms of the Ernst potential.
The latter for  the static Schwarzschild metric, with $\omega=0$ and in prolate spheroidal coordinates,  is a real function and it is given by
\begin{equation}
    \mathcal{E}=e^{2\psi}\frac{x-1}{x+1},
    \label{ES}
\end{equation}
where $\psi(x,y)$ is a real potential. By substituting the above expression into the equation of motion (\ref{eqe}), we we find that  $\psi$ satisfies
Laplace equation
\begin{equation}
    \nabla^2\psi=0.
\end{equation}
Therefore, the solution for $\psi(x,y)$ can then be written as a multipole expansion 
\begin{equation}
    \psi = \sum_{\ell\geq1} U_{\ell}(x) 
    Y_{\ell}(y) \ ,
\end{equation}
where $U_\ell$ and $Y_\ell$
satisfy 
\begin{eqnarray}
    &&\frac{\td}{\td x}\left((x^2-1)\frac{\td}{\td x}U_\ell\right)-\ell(\ell+1)U_\ell=0, \label{Xl}\\
    &&\frac{\td}{\td y}\left((1-y^2)\frac{\td}{\td y}Y_\ell\right)+\ell(\ell+1)Y_\ell=0. \label{Yl}
\end{eqnarray}
The regular solution of Eq. (\ref{Yl}) at 
$y=\pm 1$    is given by the Legendre polynomials 
\begin{eqnarray}
    Y_\ell(y)=P_\ell(y), \qquad \ell=0,1,\cdots, 
\end{eqnarray}
and similarly, the solution to Eq. (\ref{Xl}) is 
\begin{eqnarray}
    U_\ell&=&\alpha_\ell \, x^\ell \, {}_2F_1\left(\frac{1-\ell}{2},-\frac{\ell}{2},\frac{1-2\ell}{2},\frac{1}{x^2}\right)+\beta_\ell \frac{1}{x^{\ell+1}}\, {}_2F_1\left(\frac{1+\ell}{2},\frac{2+\ell}{2},\frac{3+2\ell}{2},\frac{1}{x^2}\right), \nonumber\\
    &&
    \label{fund}
\end{eqnarray}
so that the function $U(x,y)$ turns out to be
\begin{eqnarray}
    \psi(x,y)&=&\sum_{\ell=0}^\infty \bigg[\alpha_\ell \, x^\ell \, {}_2F_1\left(\frac{1\!-\!\ell}{2},-\frac{\ell}{2},\frac{1\!-\!2\ell}{2},\frac{1}{x^2}\right)+\frac{\beta_\ell}{x^{\ell+1}}\, {}_2F_1\left(\frac{1\!+\!\ell}{2},\frac{2\!+\!\ell}{2},\frac{3\!+\!2\ell}{2},\frac{1}{x^2}\right)\bigg] P_\ell(y). \nonumber\\
    &&
    \label{Uf}
\end{eqnarray}
We have seen in Ref. ~\cite{Kehagias:2024rtz}, that the decaying mode (proportional to $r^{-\ell-1}$) generates a naked singularity at the horizon $x=1$. On the other hand, the growing mode (proportional to $r^\ell$) is not singular at the horizon. Therefore, $\beta_\ell=0$ which leads to the vanishing of the static Love number for Schwarzschild BH in an external gravitational field at all orders in the tidal parameter~\cite{Kehagias:2024rtz}.  

Let us also note that 
the solution in Eq. (\ref{Uf}) determines also the function $\gamma(x,y)=\gamma_{s}(x,y)$ for the Schwarzschild BH by the equations 
(\ref{gamma}), which now are written explicitly
in  prolate spheroidal coordinates as 
\begin{align}
    \gamma_{s,x} &= \frac{1 - y^2}{x^2 - y^2} \bigg[ x \left(x^2 - 1\right) \psi_{,x}^2 - x \left(1 - y^2\right) \psi_{,y}^2 - 2y \left(x^2 - 1\right) \psi_{,x} \psi_{,y} \bigg] ,\nonumber\\
    \gamma_{s,y} &= \frac{x^2 - 1}{x^2 - y^2} \bigg[ y \left(x^2 - 1\right) \psi_{,x}^2 - y \left(1 - y^2\right) \psi_{,y}^2 + 2x \left(1 - y^2\right) \psi_{,x} \psi_{,y} \bigg]. 
\end{align}
Then, the general solution for
$\gamma_{s}(x,y)$, is provided by the closed  formula~\cite{Zipoy:1966btu}
\[\gamma_{s}(x, y) = (x^2 - 1) \int_{-1}^y \frac{\Gamma(x, y')}{x^2 - {y'}^2} \, {\rm d}y' ,\] 
where 
\[\Gamma(x, y) = y (x^2 - 1) \psi_{,x}^2 - y (1 - y^2) \psi_{,y}^2 + 2x (1 - y^2) \psi_{,x} \psi_{,y}\ .\]

\section{ Kerr BH in external tidal fields }

In order to introduce rotation, one needs to consider non-zero $\omega$ in the metric (\ref{metric}). In this case, we expect (\ref{metric}) to describe the Kerr BH as well as its embedding in external tidal fields, much the same way as in the non-rotating Schwarzschild background we described in the previous section.  
Since for a rotating BH $\omega$ is not vanishing, the Ernst potential should have a non-zero  imaginary part $\phi$, which is 
is determined by Eq. (\ref{phi}). 
 
In particular, it has been  shown~\cite{BRETON19977,TOMIMATSU1984374} that the correct choice for the Ernst potential  for a Kerr BH in an external tidal gravitational  field has the form
\begin{equation}
    \mathcal{E} = e^{2 \psi} \, \frac{x(1 + ab) + i y (b - a) - (1 - ia)(1 - ib)}{x(1 + ab) + i y (b - a) + (1 - ia)(1 - ib)}  , \label{EK}
\end{equation}
where $a=a(x,y)$ and $b=b(x,y)$. 
Then the equations of motion (\ref{eqe}) for $\mathcal{E}$ turn out to be following equations for $a(x,y)$, $b(x,y)$ and $\psi(x,y)$ 
\begin{align}
    \nabla^2\psi&=0,\nonumber\\
    (x - y) a_{,x} &= 2a \left[ (xy - 1) \psi_{,x} + (1 - y^2) \psi_{,y} \right] ,\nonumber\\
    (x - y) a_{,y} &= 2a \left[ - (x^2 - 1) \psi_{,x} + (xy - 1) \psi_{,y} \right] , \label{psiab}\\
    (x + y) b_{,x} &= -2b \left[ (xy + 1) \psi_{,x} + (1 - y^2) \psi_{,y} \right] ,\nonumber\\
    (x + y) b_{,y} &= -2b \left[ - (x^2 - 1) \psi_{,x} + (xy + 1) \psi_{,y} \right].\nonumber
\end{align}
In addition, Eq. (\ref{phi})  is written explicitly as
\begin{align}
    \mathcal{\phi}_{,x} &= \rho_0^{-1} (x^2 - 1)^{-1} f^2 \omega_{,y}\ ,\nonumber\\ 
    \mathcal{\phi}_{,y} &= \rho_0^{-1} (y^2 - 1)^{-1} f^2 \omega_{,x}.  \label{phix}
\end{align}
Then, the functions $f$, $\gamma$ and $\omega$ in the metric (\ref{metric}) turn out to be :
\begin{align}
    f &= e^{2\psi} AB^{-1},\nonumber\\
    \quad e^{2\gamma} &= K_1 (x^2 - 1)^{-1} e^{2\gamma_s} A,\\
    \omega &= 2\rho_0 e^{-2\psi} A^{-1} C + K_2,\nonumber 
\end{align}
where
\begin{align}
    A &= (x^2 - 1)(1 + ab)^2 - (1 - y^2)(b - a)^2 ,\nonumber \\ 
    B &= \left[ x + 1 + (x - 1) ab \right]^2 + \left[ (1 + y) a + (1 - y) b \right]^2 ,\nonumber \\
    C &= (x^2 - 1)(1 + ab) \left[ b - a - y (a + b) \right] + (1 - y^2)(b - a) \left[ 1 + ab + x (1 - ab) \right].  \label{ABC}
\end{align}
In Eq. (\ref{ABC}),  $K_1$ and $K_2$ are constants, whereas $\gamma_s$ is the potential $\gamma$ of the corresponding static metric with $U=\frac{1}{2}\ln\left(\frac{x-1}{x+1}\right)+\psi$.

\subsection{The Kerr  metric in Weyl  coordinates}
 
For  $a=b=0$, the Ernst potential in Eq.(\ref{EK}) reduces to the corresponding potential of Eq.  (\ref{ES}) for the Schwarzschild BH. We will in the following demonstrate, similarly, we can recover the Kerr metric from the potential in Eq. (\ref{EK}).  This is possible   when
\begin{eqnarray}
    a=-\alpha ,\qquad \ b=\alpha, \qquad \alpha=\mbox{const.}. \label{abK}
\end{eqnarray}
In this case we find that ~\cite{Castejon-Amenedo:1990yso}
\begin{align}
    {\rm Re}\{\mathcal{E}\}\equiv f =& \frac{p^2 x^2 + q^2 y^2 - 1}{(p x + 1)^2 + q^2 y^2},\nonumber\\
    e^{2\gamma}=&\frac{(p x)^2+(q y)^2-1}{p^2(x^2-y^2)},
    \nonumber\\
    \omega=&-2\rho_0\frac{q\left(px+1 \right)\left(1-y^2\right)}{p(p^2 x^2 + q^2 y^2 - 1)},
\end{align}
where,
\begin{equation}
    p = \frac{1 - \alpha^2}{1 + \alpha^2}, \quad q = \frac{2\alpha}{1 + \alpha^2}, \quad p^2 + q^2 = 1.
\end{equation}
In the same spirit, we can substitute the imaginary part of Ernst potential in Eq. (\ref{phix}) to find $\omega$ and 
by using Eqs. (\ref{ABC}) with 
\begin{equation}
    K_1 = \frac{1}{(1 - \alpha^2)^{2}}, \qquad K_2 = -\frac{4\rho_0 \alpha}{1 - \alpha^2},\label{k1k2}
\end{equation} 
we end up with  Kerr metric  in prolate coordinates. The transition to Boyer–Lindquist coordinates can be made by the following  set of substitutions
\begin{equation}
    \rho_0x = r - m,\quad y=\cos\theta ,\quad \rho_0 = mp, \quad a_0 = mq, \quad \rho_0^2 = m^2 - a_0^2 ,\label{boyer1}
\end{equation}
where $m$ is the BH mass and $a_0$ is the spin parameter of the Kerr BH. In addition, since the spin always satisfy $m^2\leq a_0^2$, it follows from Eq. (\ref{boyer1}) that the range of $\alpha$ is $|\alpha|\leq 1$. We therefore end up with the known form of the Kerr metric:
\begin{align}
     {\rm d}s^2 &= -\left(1 - \frac{2m r}{\Sigma}\right) {\rm d}t^2 + \frac{\Sigma}{\Delta} {\rm d}r^2 + \Sigma {\rm d}\theta^2 - \frac{4 m  a_0 r \sin^2 \theta}{\Sigma} \, {\rm d}t \, {\rm d}\varphi\nonumber\\
    &\quad + \left(r^2 + a_0^2 + \frac{2m  a_0^2r}{\Sigma} \sin^2 \theta\right) \sin^2 \theta {\rm d}\varphi^2 \ ,\label{KBL}
\end{align}
where, as usual,
\begin{eqnarray}
    \Delta=r^2-2m r+a_0^2, \qquad 
    \Sigma=r^2+a_0^2 \cos^2\theta. 
\end{eqnarray}
Therefore, we see that indeed, the Ernst potential (\ref{EK}) with $a$ and $b$ as in Eq. (\ref{abK}) describes the Kerr metric in prolate spheroidal coordinates.\\

While Boyer-Lindquist coordinates are preferable in general for describing the Kerr metric some tasks that we encounter later in this paper  seem to prefer treatment using  Weyl spherical coordinates. Spherical coordinates $(R,u,\varphi)$ can be expressed in terms of Weyl canonical coordinates $(\rho,z,\varphi)$ and Boyer-Lindquist coordinates $(r,\theta,\varphi)$ as
    \begin{align}
    R=&\sqrt{\rho^2+z^2}=\sqrt{(r-m)^2-\rho_0^2\sin^2\theta} \ ,\nonumber\\
    \cos u=&\frac{z}{\sqrt{\rho^2+z^2}}=\frac{(r-m)\cos \theta}{\sqrt{(r-m)^2-\rho_0^2\sin^2\theta}} . \label{Ru}
\end{align}
Transitions between all the previously mentioned coordinate systems can be significantly simplified using the auxiliary functions $R_+$ and $R_-$ defined as 
\begin{equation}
    R_\pm(\rho,z)=\sqrt{\rho^2+(z\pm\rho_0)^2}=(r-m)\pm\rho_0\cos\theta=\sqrt{R^2+\rho_0^2\pm2\rho_0 R  \cos u} \ . \label{Rpm}
\end{equation}
Note that, with the use of (\ref{boyer1}) one obtains quite trivially the inverse transformation of Eq. (\ref{pro}) as 
\begin{align}
    \rho_0x&=\frac{1}{2}\left(R_+ + R_- \right)=\frac{1}{2}\left( \sqrt{\rho^2+(z+\rho_0)^2}+\sqrt{\rho^2+(z-\rho_0)^2}\right),\nonumber\\
    \rho_0y&=\frac{1}{2}\left(R_+ - R_- \right)=\frac{1}{2}\left(\sqrt{\rho^2+(z+\rho_0)^2}-\sqrt{\rho^2+(z-\rho_0)^2} \right), \label{tr1}
\end{align}
In the new coordinates $(R,u,\varphi)$, the Kerr metric is written as
\begin{equation}
    {\rm d}s^2 = f^{-1} \left[ e^{2\gamma} ({\rm d}R^2 + R^2 {\rm d}u^2) + R^2\sin^2u\ {\rm d}\varphi^2 \right] - f ({\rm d}t -\omega {\rm d}\varphi)^2, \label{rr1}
\end{equation}
where
\begin{align}
    f&= 1-\frac{ 4m(R_+ + R_-+2m) }{(R_+ + R_-+2m)^2 + \frac{a_0^2}{m^2 - a_0^2} (R_+ - R_-)^2} ,\nonumber\\
    e^{2\gamma} &= \frac{(R_+ + R_-)^2 - 4m^2 + \frac{a_0^2}{m^2 - a_0^2}(R_+ - R_-)^2}{4R_+ R_-},\\
    \omega&=- \frac{a_0m( R_+ + R_- +2m)(4 - \frac{(R_+ - R_-)^2}{(m^2 - a_0^2)})}{(R_+ + R_-)^2 - 4m^2 + a_0^2\frac{(R_+ - R_-)^2}{(m^2 - a_0^2)}}. \nonumber
\end{align}
Finally, the metric (\ref{rr1}) can be rewritten in the known form of (\ref{KBL}) in terms of the  Boyer-Lindquist coordinates $(r,\theta,\varphi)$ by the coordinate transformations of (\ref{Ru}) and (\ref{Rpm}).

\section{The Kerr BH in external tidal fields}

An inspection of Eqs. (\ref{psiab}) shows that both $a$ and $b$ are determined only up to a multiplicative constant. Therefore, we can utilize this freedom by choosing always the constant value of $a$ and $b$ as in Eq. (\ref{abK}).  
Let us now rewrite the metric in Eq. (\ref{metricx}) in the following way 
\begin{equation}
    {\rm d}s^2 = - f \left( {\rm d}t - \omega {\rm d}\varphi \right)^2 + h \left( \frac{{\rm d}x^2}{x^2 - 1} + \frac{{\rm d}y^2}{1 - y^2} \right) 
+ \rho_0^2 f^{-1} (x^2 - 1)(1 - y^2) {\rm d}\varphi^2\ , \label{met2}
\end{equation}
where
\begin{align}
    h&=\frac{\rho_0^2}{(1 - \alpha^2)^2} B e^{-2\psi + 2V} ,\\
    V&=\gamma_s-\frac{1}{2}\ln \left(\frac{x^2-1}{x^2-y^2} \right) .
\end{align}
Form Eqs. (\ref{psiab}) we see that    $\psi$ satisfies the Laplace equation and thus a general solution in Weyl spherical harmonics must be of the form
\begin{equation}
    \psi=\sum_{\ell\geq1}\left(c_\ell R^{\ell}+ \frac{d_{\ell}}{R^{\ell+1}}\right)P_{\ell}( \cos u ), \label{psiR}
\end{equation}
where 
$R=R(x,y)$ and $u=u(x,y)$. 
Notice that the series starts from $\ell=1$ since the $\ell=0$ term yields the Kerr solution and has been factored out in the parametrization of the Ernst potential in Eq. (\ref{EK}). 
It also important to point out that in order to obtain well-defined solutions without conical singularities along the symmetry axis one has to consider the following condition~\cite{xanthopoulos1983local}
\begin{align}
    \sum_{n=0}^{\infty} c_{2n+1} = 0\ .
\end{align}
Therefore, we cannot have a single dipole  without an octupole tidal deformation.

\subsection{The decaying quadrupole mode}

The solution for $\psi$ in Eq. (\ref{psiR})  is the sum of  decaying modes (proportional to $R^{-\ell-1}$) and growing modes (proportional to $R^\ell)$. Here, we will examine the quadrupole modes, consequently the solution for $\psi$ that we will consider will be of the form:
\begin{equation}
    \psi = \left( c_2 R^2 + \frac{d_2}{R^3} \right)P_2( \cos u ),  \label{psiRR}
\end{equation}
where $c_2$ and $d_2$ are the strength of the growing and decaying tidal fields, respectively.  
We can now calculate the general expressions for $a$, $b$, and $V$ using equations (2.13) and (2.15) for $\gamma_s$ 
\allowdisplaybreaks
\begin{align}
    a(x,y)=&-\alpha \exp\bigg\{2 c_2 (x y +1)(x-y) -d_2\left[\left(x^2+y^2-1\right)^{-5/2} \left(2 x^5+5 x^3 \left(y^2-1\right)\right.\right.
    \nonumber\\ &\left.\left. -x^2 y \left(5 y^2-3\right)-3 x \left(y^2-1\right)-y \left(2 y^4-5 y^2+3\right)\right)-2\right]\bigg\},\\
    b(x,y)=&\alpha \exp\bigg\{2 c_2 (1-xy)(x+y)-d_2\left[\left(x^2+y^2-1\right)^{-5/2}\left(2 x^5+5 x^3 \left(y^2-1\right)\right.\right.
    \nonumber\\ &\left.\left.+x^2 y \left(5 y^2-3\right)-3 x \left(y^2-1\right)+y \left(2 y^4-5 y^2+3\right)\right)+2\right]\bigg\},\\
    V(x,y)=&-\frac{1}{8} \left(y^2-1\right) \left(-2 c_2^2 x^4 \left(9 y^2-1\right)+4 c_2^2 x^2 \left(5 y^2-1\right)-\frac{24 d_2 y^2 \left(-c_2 y^4+c_2 y^2+x\right)}{\left(x^2+y^2-1\right)^{5/2}}+\right.
    \nonumber\\ &\left.\frac{8 d_2 \left(x-6 c_2 y^4\right)}{\left(x^2+y^2-1\right)^{3/2}} +16 c_2 x+\frac{75 d_2^2 \left(y^2-1\right)^2 y^6}{\left(x^2+y^2-1\right)^6}-\frac{9 d_2^2 \left(25 y^4-38 y^2+13\right) y^4}{\left(x^2+y^2-1\right)^5}\right.\nonumber\\& \left.+\frac{9 d_2^2 \left(25 y^4-26 y^2+5\right) y^2}{\left(x^2+y^2-1\right)^4}-\frac{3 d_2^2 \left(25 y^4-14 y^2+1\right)}{\left(x^2+y^2-1\right)^3}\right)+\frac{c_2^2 y^4}{4}-\frac{c_2^2 y^2}{2}\nonumber\\&+\frac{d_2 \left(3 c_2 \left(y^4-1\right)+2 x\right)}{\sqrt{x^2+y^2-1}}+V_0. \label{V0}
\end{align}
The constant $V_0$ in Eq. (\ref{V0}) is determined by the  regularity condition~\cite{xanthopoulos1983local} $ \lim_{y\to \pm 1}\gamma(x,y)=0 $. 
Since $d_2$ in Eq. (\ref{psiRR}) is proportional to the static TLN for quadrupole tidal deformations, we will consider below only the decaying mode. Then, with  $c_2=0$, we find that 
\allowdisplaybreaks
\begin{align}
    \psi(x,y)=&d_2\frac{1}{R^3}P_2(\cos u),\\
    a(x,y)=&- \alpha\exp\bigg\{-d_2\Big[w(x,y)-w(y,x)\Big]\bigg\},\label{decpsi}\\
    b(x,y)=& \alpha\exp\bigg\{-d_2\Big[w(x,y)+w(y,x)\Big]\bigg\},\\
    A(x,y)=&4\alpha^2\exp\bigg\{-2d_2w(x,y)\bigg\}\bigg\{(x^2-1)\sinh^2\Big(d_2w(x,y)-\ln |\alpha| \Big)\nonumber\\
    & + (y^2-1)\cosh^2\Big(d_2w(y,x)\Big)\bigg\},\\   
    B(x,y)=&4\alpha^2\exp\bigg\{-2d_2w(x,y)\bigg\}\bigg\{\left(\sinh\Big(d_2w(x,y)-\ln|\alpha|\Big)\right.\nonumber\\
    &\left.+\cosh\Big(d_2w(x,y)-\ln\abs{\alpha}\Big)\right)^2+\left(y\cosh\Big(d_2w(y,x)\Big) +\sinh\Big(d_2w(y,x)\Big)\right)^2 \bigg\},\\  
    C(x,y)=&4\,{\rm sign}\alpha\,\alpha^2\exp\bigg\{-2d_2w(x,y)\bigg\}\bigg\{(x^2-1)\sinh\Big(d_2w(x,y)-\ln|\alpha|\Big)\left(\cosh\Big(d_2w(y,x)\Big)\right.\nonumber\\
    &\left.+y\sinh\Big(d_2w(y,x)\Big)\right)+(1-y^2)\cosh\Big(d_2w(y,x)\Big)\left( \sinh\Big(d_2w(x,y)-\ln|\alpha|\Big)\right.\nonumber\\
    &\left.+x\cosh\Big(d_2w(x,y)-\ln|\alpha|\Big)\right) \bigg\},\label{decC}
\end{align}
where we have redefined the constant $\alpha$ as 
$\alpha e^{2d_2}$ and 
\begin{align}
    w(x,y)&=\frac{l(x,y)}{R^5(x,y)}, \nonumber
    \\
    l(x,y)&=x\Big( 2x^4 + (5x^2-3)(y^2-1) \Big)\ . \nonumber
\end{align}
 In order to keep the same notation while also including $a<0$ one can introduce complicated expressions involving the $sign\{\alpha\}$ function, but that turns out to be unnecessary for the purposes of this proof. We are now interested in determining the possible singularities, which in principle can be generated by turning on tidal fields. A way of determining the existence of singularities is by checking if curvature scalars, such as the  Kretschmann  scalar
\begin{eqnarray} \mathcal{K}=R_{\mu\nu\rho\sigma}R^{\mu\nu\rho\sigma}, \label{krei}
\end{eqnarray}
become singular. Since we are discussing the case of the gravitational response of a Kerr BH to an outer gravitational field, we would expect 
no other singularities other than the known singularities of the Kerr BH. In the opposite case, where new singularities emerge, these should be dressed with a horizon, and they cannot be naked.  Therefore, if the Kretschman scalar for example becomes singular somewhere else other than the known Kerr singularities, then the static Love numbers should vanish, provided the new singularities are naked.  
The complete expression of the  Kretschmann  scalar is quite long and not at all illuminating. However, expanding at the equator $(y=0)$ and close to the outer horizon $(x=1)$ of the Kerr BH, we find that the Kretschmann  scalar is 
\begin{align}
    \mathcal{K}(s,0)\sim \mathcal{K}_0+e^{
    \frac{3 d_2 ^2}{16 s^3}}\left(\frac{d_2^6}{s^{12}}
    + \mathcal{O}(s^{-11}) \right), \qquad s=x-1,
\end{align}
where $\mathcal{K}_0$ is the Kretschmann  scalar for the Kerr metric at $(x=1,y=0)$.
We see that  if $d_2\neq0$, then the  Kretschmann 
scalar becomes singular as $s\to0$ indicating the appearance of a naked singularity. The   singularity should be removed since it is naked. This is achieved by taking $d_2=0$, and  therefore, the Love number of the Kerr BH vanishes to any order in the tidal field. 

\subsection{The growing quadrupole mode}

We have seen above that the decaying mode leads to curvature naked singularities and therefore, the TLNs of the Kerr BH should be zero at the full non-linear level. In the following, we will similarly study the growing quadrupole $(\ell=2)$ mode. It has been shown previously~\cite{TOMIMATSU1984374} that if we keep only growing modes (as we will in our case since TLN's vanish), then analytic expressions for $a\big(R(x,y),u(x,y)\big)$ and $b\big(R(x,y),u(x,y)\big)$ can be calculated, and hence the metric components can be written explicitly in terms of Legendre polynomials for arbitrary $\ell$ as follows~\cite{BRETON19977}
\allowdisplaybreaks
\begin{align}
     \psi &= \sum_{\ell=1}^{\infty} c_\ell \left(\frac{R}{\rho_0}\right)^\ell P_\ell ( \cos u ), \\
    a =& -\alpha \exp \bigg\{ 2 \sum_{n=1}^{\infty} c_n \frac{R_-}{\rho_0} \sum_{\ell=0}^{n-1} \left(\frac{R}{\rho_0}\right)^\ell P_\ell ( \cos u  )\bigg\} ,\\
    b =& \alpha \exp \bigg\{ 2 \sum_{n=1}^{\infty} c_n \frac{R_+}{\rho_0} \sum_{\ell=0}^{n-1} (-1)^{n-\ell} \left(\frac{R}{\rho_0}\right)^\ell P_\ell ( \cos u  )\bigg\},\\ 
    V =& \sum_{\ell,m=1}^{\infty} \frac{\ell m}{\ell + m} c_\ell c_m \left(\frac{R}{\rho_0}\right)^{\ell+m} \bigg[P_\ell P_m - P_{\ell-1} P_{m-1}\bigg] \nonumber \\
    &
    + \sum_{\ell=1}^{\infty} c_\ell \sum_{m=0}^{\ell-1} \left[ (-1)^{\ell-m+1} \frac{R_+}{\rho_0} - \frac{R_-}{\rho_0}\right] \left(\frac{R}{\rho_0}\right)^m P_m,\\
    h=&\frac{\rho_0^2}{(1-a^2)^2}Be^{2(V-\psi)},\\
    \omega =& 2 \rho_0 e^{-2\psi} \frac{C}{A} - \frac{4 \rho_0 \alpha}{1 - \alpha^2} \exp \left( -2 \sum_{n=0}^{\infty} c_{2n} \right).
\end{align}
For the quadrupole $\ell=2$ deformations we are interested in,  the  potentials $\psi$, $\gamma_s$  and $V$  are then written 
\begin{align}
    \psi=&c_2\left(\frac{R}{\rho_0}\right)^2P_2(\cos u),\\
    \gamma_s=&\frac{1}{2}\ln\left(\frac{(R_++R_-)^2-4\rho_0^2}{4R_+R_-} \right)+ c_2 ^2 \left(\frac{R}{\rho_0} \right)^4\Big(P_2^2(\cos u)-P_1^2(\cos u)\Big)\nonumber\\&+c_2\left( \frac{R_+}{\rho_0}\left(\frac{R}{\rho_0}\cos u-1\right) -\frac{R_-}{\rho_0}\left(\frac{R}{\rho_0}\cos u+1\right)\right),\nonumber\\
    V=&  c_2 ^2 \left(\frac{R}{\rho_0} \right)^4\Big(P_2^2(\cos u)-P_1^2(\cos u)\Big)+c_2\left( \frac{R_+}{\rho_0}\left(\frac{R}{\rho_0}\cos u-1\right) -\frac{R_-}{\rho_0}\left(\frac{R}{\rho_0}\cos u+1\right)\right),\nonumber\\
\end{align}
and therefore, we find 
\allowdisplaybreaks
\begin{align}
   f=&e^{2 \psi}\frac{((R_+ + R_-)^2 - 4\rho_0^2)(1 + ab)^2 - (4\rho_0^2 - (R_+ - R_-)^2)(b - a)^2}{\left[ (R_+ + R_-)(1+ab)+2\rho_0(1-ab) \right]^2 + \left[ 2\rho_0(a+b)+(R_+ - R_-)(a-b)\right]^2},\\
   f^{-1}e^{2\gamma}=&\frac{e^{2(\gamma_s-\psi)}}{(1-a^2)^2}\frac{\left[ (R_+ + R_-)(1+ab)+2\rho_0(1-ab) \right]^2 + \left[ 2\rho_0(a+b)+(R_+ - R_-)(a-b)\right]^2}{(R_+ + R_-)^2-4\rho_0^2},\\
   \omega=&e^{-2\psi}\frac{((R_+ + R_-)^2 - 4\rho_0^2)(1 + ab)(2\rho_0(b-a)-(R_+-R_-)(a+b))}{((R_+ + R_-)^2 - 4\rho_0^2)(1 + ab)^2 - (4\rho_0^2 - (R_+ - R_-)^2)(b - a)^2}+\label{ww}\\
   &+\frac{ (4\rho_0^2 - (R_+ - R_-)^2)(b - a)(2\rho_0(1+ab)+(R_++R_-)(1-ab))}{((R_+ + R_-)^2 - 4\rho_0^2)(1 + ab)^2 - (4\rho_0^2 - (R_+ - R_-)^2)(b - a)^2}-\frac{4\rho_0\alpha}{1-\alpha^2}e^{-2 c_2} ,\nonumber 
\end{align}
where
\begin{align}
    a=&-\alpha \exp\bigg\{2 c_2 \frac{R_-}{\rho_0}\bigg[1+\frac{R}{\rho_0}\cos u\bigg]\bigg\} ,\nonumber \\
    b=&\alpha \exp\bigg\{2 c_2 \frac{R_+}{\rho_0}\bigg[1-\frac{R}{\rho_0}\cos u\bigg]\bigg\}.
\end{align}
We should now examine if there are also naked singularities for the growing mode as well. 
It has been shown in~\cite{TOMIMATSU1984374}, that for $x>1$ singularities arise whenever $B=0$, where $B$ has been defined in Eq. (\ref{ABC}). By assuming that  $c_2<0$,  we find that $B\neq 0$ and therefore, there are no singularities in $x>1$ in this case. So, the only possibility is to have singularities on the horizon at $x=1$. Similarly to the decaying mode, the calculation of the  Kretschmann scalar $\mathcal{K}(x,y)$ around  $x=1$  shows that 
\begin{align}
    \mathcal{K}(1,y)=\frac{1}{ \left(\alpha ^2 y^2+e^{4 c_2 \left(y^2-1\right)}\right)^6} P(y), \label{KP1}
\end{align}
where $P(y)$ is a polynomial in $y$. 
By examining the expression (\ref{KP1}) analytically we realize that for $c_2<0$ and $y\in[-1,1]$ the denominator is non zero, therefore no singularities arise at the outer horizon of the BH. The behavior (\ref{KP1}) around $x=1$, is in accordance with the plots of the Kretschmann scalar given in ~\cite{Abdolrahimi:2015gea}. It is also important to highlight that the physical consequence of $c_2<0$ is that the Kerr BH slows down its rotation when tidal fields are present. This can be formally understood by calculating the angular velocity at the horizon of the BH. One obtains the angular velocity expression in Boyer-Lindquist coordinates as ~\cite{Townsend:1997ku} 
\begin{align}
    \Omega_H=-\frac{g_{tt}}{g_{t\phi}}\bigg|_H .\label{omegaH}
\end{align}
By substituting our findings for the quadrupole in (\ref{omegaH}) we obtain
\begin{align}
    \Omega_H=\frac{a_0 }{a_0^2+r_+^2}e^{2 c_2}=
    \Omega_H^Ke^{2 c_2}, \label{omegaH1}
\end{align}
where $ \Omega_H^K=a_0/a_0^2+r_+^2$ is the angular velocity of the horizon of the Kerr metric. Therefore, as Eq. (\ref{omegaH1}) indicates, a Kerr BH would spin up for $c_2>0$, which is not physically plausible. In reality, due to tidal braking, a Kerr BH should slow down its rotation, leading to a reduction in its angular momentum when subjected to a tidal field. 
This behavior is consistent only if 
 $c_2<0$, which also explains the emergence of singularities for $c_2>0$.

\section{The role of symmetries}

We have seen above that the static tidal Love numbers  vanish identically at the full non-linear level  not only for a non-rotating BH~\cite{Kehagias:2024rtz,Combaluzier-Szteinsznaider:2024sgb}, but also for rotating BHs, suggesting that there is an underlying non-linear symmetry explaining such a behavior also in the case of rotating spacetimes. Such a  symmetry already appears   at  the linear level in  the tidal force \cite{Hui:2020xxx,Charalambous:2021mea,Charalambous:2021kcz,Hui:2021vcv,Hui:2022vbh,Charalambous:2022rre,Ivanov:2022qqt,Katagiri:2022vyz, Bonelli:2021uvf,Kehagias:2022ndy,BenAchour:2022uqo,Berens:2022ebl,DeLuca:2023mio, Rai:2024lho}. In fact, it turns out that for each  mode $\ell$ solving these equations, a conserved quantity $P_\ell$ exists which is associated with the aforementioned underlying symmetry. The corresponding conserved charges allow for descending to the monopole case ($\ell = 0$) using ladder operators. Conservation of $P_0$ implies the invariance of $P_\ell$ for higher modes, providing a framework to understand why the decaying solution $\sim 1/r^{\ell+1}$ must be excluded, as it is tied to divergences at the horizon. The non-linear version of the symmetry has been identified in Refs. \cite{Kehagias:2024rtz,Combaluzier-Szteinsznaider:2024sgb}.

Now, a pivotal observation is that the equation for $\psi$  which governs the static configuration even in the full non-linear regime, retains a linear structure as it solves the Laplace equation. Remarkably, this equation coincides with the one solved in the linear case for a static, massless scalar field in the Schwarzschild background. However, the non-linearities here are encoded in the function $a(x, y)$ and $b(x,y)$, which enter the parametrization of the Ernst potential in Eq. (\ref{EK}). Expanding $\psi(x,y)$ as~\cite{Kehagias:2024rtz}
\begin{eqnarray}
    \psi(x,y)=\sum_{\ell=0}U_\ell(x) P_\ell(y),
\end{eqnarray}
we can 
define the following ladder operators as
\begin{align}
    L^+_\ell &= -(x^2 - 1)\frac{\td}{\td x} - (\ell + 1)x, \nonumber \\
    L^-_\ell &= (x^2 - 1)\frac{\td}{\td x} - \ell x.
    \label{ladder_ops}
\end{align}
These operators act as raising and lowering operators for the multipole moments, satisfying
\begin{equation}
    L^+_\ell U_\ell \sim U_{\ell+1}, \quad L^-_\ell U_\ell \sim U_{\ell-1}.
\end{equation}
Then following the standard constructions used in linear perturbation theory \cite{Hui:2021vcv}, one can define conserved quantities
\begin{equation}
    P_\ell = (x^2 - 1)\frac{\td}{\td x} \left( L^-_1 L^-_2 \cdots L^-_\ell \right)U_\ell,
\end{equation}
for which:
\begin{equation}
    \frac{\td P_\ell}{\td x} = 0.
\end{equation}
For the decaying solution, we find that at large $x$
\begin{equation}
    U_\ell \sim \frac{\beta_\ell}{x^{\ell+1}},
\end{equation}
resulting in a conserved $P_\ell$ that remains finite but non-zero as $x \to \infty$. Near the event horizon, this decaying mode diverges logarithmically, as $\ln(x-1)$. Since the growing and decaying modes must share the same $P_\ell$, and the growing mode at the horizon is constant (implying $P_\ell = 0$), the conservation of $P_\ell$ necessitates the exclusion of the decaying solution due to its divergence.
However, an additional argument is required. The reason is that, due to the aforementioned divergence, linear perturbation theory breaks down, and one has to consider the full non-linear problem. We found here that indeed the divergence of the decaying mode at the horizon survives at the full non-linear lever and shows off as a naked singularity as the Kretschmann scalar indicates. Therefore, the decaying mode should be completely eliminated leading to a vanishing static Love number. By discarding the decaying modes in $\psi(x,y)$, no extra divergences propagate into the non-linear Ernst potential, ensuring consistency with the Kerr background.

The Laplacian equation satisfied by $\psi$ in Eq. (\ref{psiab}) is structurally equivalent to that in a two-dimensional flat spacetime in the original Weyl coordinates $(\rho,z)$. Its solutions can therefore be expressed in terms of holomorphic functions
\begin{equation}
    \psi(\zeta, \bar{\zeta}) = \varPsi(\zeta) + \bar{\varPsi}(\bar{\zeta}),
\end{equation}
where $\zeta = \rho + iz$. Any analytic transformation of $\zeta$ yields a new solution. Then the ladder operators are generators of a conformal symmetry group associated with these holomorphic transformations.

We should note however, that the above conformal (homolorphic) symmetries are tied up to the symmetries of the Ernst potential $\mathcal{E}$. 
An inspection of the action (\ref{ee}) or of Eq. (\ref{eqe}) reveals that they are both
invariant under the SL$(2,\mathbb{R})$ group which act on the Ernst potential as 
\begin{equation}
  \mathcal{E}\to    \mathcal{E}'=-i\frac{ai\mathcal{E}+b}{ci\mathcal{E}+d}, \qquad ad-bc=1,
\end{equation}
or in terms of $f$ and $\phi$ 
\begin{align}
 \phi\to    \phi'=&-\frac{ac f^2+(d-c\phi) (b-a \phi) }{c^2 f^2 +(d-c \phi)^2}\nonumber \\
    f\to f'= &\frac{f }{c^2 f^2 +(d-c \phi)^2}.
    \label{sl}
\end{align}
The action  (\ref{ee}) and the equations (\ref{eqe}) are identical to the action and the equations of motion of a non-linear SL$(2, \mathbb{R}) / $U(1) $\sigma$-model in two dimensions. This can be seen from the parametrization of the SL$(2,\mathbb{R})$ group 
by the $2\times 2$ matrices 

\begin{eqnarray}
    V=\left(\begin{array}{cc}
        V_-^1&V_+^1\\
        V_-^2&V_+^2
    \end{array}\right)=\frac{i}{\sqrt{-2i f}}
    \left(\begin{array}{cc}
        -\mathcal{E}^*e^{-i\vartheta}&\mathcal{E}e^{i\vartheta}\\
        \mathcal{E}^*e^{-i\vartheta}&\mathcal{E}^*e^{i\vartheta}
        \end{array}\right).
\end{eqnarray}
There is a local U(1) which is realized by the shifts $\vartheta\to \vartheta +\Delta\vartheta$, and a global SL$(2, \mathbb{R})$ that acts from the
left. Clearly then, $\mathcal{E}$ parameterizes the SL$(2,\mathbb{R})/$U(1) coset space once the local U(1) is fixed.  
 Such non-linear $\sigma$-models appear frequently in GR and are often referred to as Ernst models. Originally introduced in the context of Geroch’s reduction of GR~\cite{Geroch:1972yt} and extensively studied by Ernst~\cite{Ernst:1967wx}, these models provide a framework to understand the symmetry properties of stationary solutions in GR.

 In fact, the {SL}(2,$\mathbb{R})$ symmetry of the Ernst model, due to mixing with the larger conformal (holomorphic)  transformations give rise to an infinite algebra, the {SL}(2,$\mathbb{R})$ infinite dimensional current algebra. The ladder operators stemming out from  the Laplace equation are indeed part of the generators of this infinite-dimensional group of transformations, which therefore explain the vanishing of the static TLN for four-dimensional BHs. 

This symmetry structure is a hallmark of stationary and axisymmetric spacetimes, which are inherently linked to the two-dimensional nature of the equations governing such systems.
The infinite-dimensional symmetry described above governs the solution space of stationary, axisymmetric spacetimes in Einstein's vacuum field equations. 
The Ernst models in two-dimensions and the associated symmetry structures have been widely used in studying BH solutions, including the generation of exact solutions such as the Kerr metric or multi-BH configurations. They are also crucial in exploring extensions of general relativity, where similar two-dimensional dynamics occur.

All of the above underscore the rich symmetry structure inherent in the two-dimensional reduction of general relativity. This structure, exemplified by the infinite-dimensional SL$(2, \mathbb{R})$ algebra, provides a powerful tool for understanding stationary, axisymmetric solutions and underlines the vanishing of the static tidal Love numbers~\cite{Kehagias:2024rtz, Combaluzier-Szteinsznaider:2024sgb}.

\section{Conclusions}
In this paper, we have analyzed the non-linear tidal response of Kerr BHs under the influence of external gravitational fields. Using the Ernst formalism and Weyl coordinates, we systematically extended previous results for Schwarzschild BHs to the case of rotating Kerr BHs. Our primary finding is that the static tidal Love numbers of Kerr BHs vanish at all orders in the external tidal field, consistent with the unique symmetries and characteristics of these spacetimes.

The vanishing of the static Love numbers reflects the absence of internal structure in BHs and the profound influence of their underlying spacetime symmetries. Unlike neutron stars, which exhibit nonzero Love numbers that depend on their internal composition, BHs are characterized by their event horizons and the no-hair theorem. This result implies that Kerr BHs cannot sustain any multipole deformations in response to external tidal forces, even when higher-order non-linear effects are taken into account. It also emphasizes the resilience of BH spacetimes against tidal perturbations, a property that distinguishes them from other compact objects.

Our analysis highlighted the utility of the Ernst potential in describing the behavior of BHs in tidal environments. By expressing the Kerr metric in Weyl coordinates, we were able to generalize the Schwarzschild case and examine the role of rotational effects. The use of prolate spheroidal coordinates further facilitated the derivation of key results, enabling a rigorous examination of both growing and decaying quadrupole modes. The identification of singularities in the Kretschmann scalar associated with the decaying mode underscores the physical consistency of setting the Love numbers to zero. This approach reaffirms that any tidal-induced singularity must remain hidden behind a horizon, preserving the integrity of the spacetime.

From an astrophysical perspective, the vanishing of Kerr BH Love numbers has significant implications for gravitational wave astronomy. The tidal deformability of BHs is a critical parameter in the modeling of waveforms from binary inspirals, particularly in scenarios involving BH-neutron star or BH-BH mergers. The lack of tidal signatures from BHs simplifies waveform modeling while providing a stringent test of general relativity in the strong-field regime. Furthermore, these results help refine the theoretical foundations for interpreting gravitational wave data, ensuring that deviations from predicted signals are not misattributed to unmodeled BH tidal effects.

Future research could explore several extensions of this work. One avenue is the inclusion of dynamical tidal effects, where time-dependent perturbations may lead to dissipative phenomena or resonances. Another is the investigation of quantum corrections to the Love numbers, particularly in contexts where semiclassical gravity or string theory might introduce additional structure to the spacetime. Finally, the study of tidal effects in higher-dimensional BHs or alternative theories of gravity could provide a broader context for understanding the universality of our findings.

In conclusion, our results reinforce the fundamental nature of BHs as geometrically simple yet profoundly enigmatic objects. The vanishing of their tidal Love numbers, even in the non-linear regime, exemplifies their remarkable symmetry and resistance to external perturbations. These findings contribute to the deeper understanding of BH physics and its pivotal role in testing the limits of general relativity.

\normalsize
\begin{acknowledgments}
\noindent
    A.K. acknowledges support from the  Swiss National Science Foundation  (project 
number  IZSEZ0\_229414). A.R.  acknowledges support from the  Swiss National Science Foundation (project number CRSII5\_213497)
and by  the Boninchi Foundation for the project ``PBHs in the Era of GW Astronomy''.
\end{acknowledgments}

\newpage
\vskip.5in
\noindent
{\bf \large Appendix}
\appendix
\section{Expansion in Boyer-Lindquist coordinates}

By expanding the Eq. (\ref{met2}) in powers of the tidal parameter $c_2$, we can find the resulting static perturbation of the Kerr metric induced by the external gravitational tidal field.
Since, the Kerr metric is usually written in  Boyer-Lindquist coordinates, it is more convenient to write the metric in these coordinates. 
We can express the general form of the metric in Eq. (\ref{met2}) in  Boyer-Lindquist coordinates by using the transformations
of Eq. (\ref{boyer1}) as 
\begin{eqnarray}
    {\rm d}s^2=g_{tt} {\rm d}t^2+2 g_{t\varphi}{\rm d}t {\rm d}\varphi+g_{rr}{\rm d}r^2
    +g_{\theta\theta}{\rm d}\theta^2
    +g_{\varphi\varphi} {\rm d}\varphi^2, 
    \label{metg}
\end{eqnarray}
where
\begin{align}
    g_{tt}&= -f(r,\theta)\ ,\nonumber\\
    g_{t\phi}&= f(r,\theta)\ \omega(r,\theta)\ ,\nonumber\\
    g_{rr}&= \frac{h(r,\theta)}{\Delta(r)}\ ,\label{metgg}\\
    g_{\theta\theta}&= h(r,\theta)\ ,\nonumber\\
    g_{\phi\phi}&= \frac{\sin^2\theta\ \Delta(r)}{f(r,\theta)}-f(r,\theta)\omega^2(r,\theta)\ .\nonumber
\end{align}
The metric in Eq. (\ref{metgg}) above can be expanded in the parameter $c_2$, which determines the strength of the tidal deformation as
\begin{align}
    &g_{ij}=g^{K}_{ij}(r,\theta) + c_2  h_{ij}(r,\theta) + \mathcal{O}(c_2^2)\ ,\label{perKerr1}
\end{align}
where 
$g^{K}_{ij}(r,\theta)$ is the Kerr metric.
The explicit form of the perturbation $h_{ij}$ is complicated and not illuminating. However, for the extreme Kerr BH ($m^2=a_0^2$), more tractable expression can be written. In this case, the extreme Kerr metric is written as
\begin{eqnarray}
    {\displaystyle {\rm d}s^{2}\,=\,-{\frac {\rho _{K}^{2}\Delta _{K}}{\Sigma_K ^{2}}}\,{\rm d}t^{2}+{\frac {\rho _{K}^{2}}{\Delta _{K}}}\,{\rm d}r^{2}+\rho _{K}^{2}{\rm d}\theta ^{2}+{\frac {\Sigma_K ^{2}\sin ^{2}\theta }{\rho _{K}^{2}}}{\big (}{\rm d}\varphi -\omega _{K}\,{\rm d}t{\big )}^{2}\,,}
    \label{kere}
\end{eqnarray}
where 
\begin{align}
     \rho _{K}^{2}=&r^{2}+m^{2}\cos ^{2}\theta ,\quad \Delta _{K}={\big (}r-m{\big )}^{2}, \quad\omega _{K}={\frac {2m^{2}r}{\Sigma_K ^{2}}}\,\nonumber \\
    \Sigma_K ^{2}=&{\big (}r^{2}+m^{2}{\big )}^{2}-m^{2}\Delta _{K}\sin ^{2}\theta. 
\end{align}
Then, we can expand the tidally deformed extreme Kerr metric in Legendre polynomials $P_\ell(\cos{\theta})$ as 
\begin{align}
    &g_{ij}=g^{K_e}_{ij} + c_2  \mathcal{H}_{ij}^\ell(r) P_\ell(\cos{\theta}) + \mathcal{O}(c_2^2), \label{perKerr2}
\end{align}
where $g^{K_e}_{ij}$ is the extreme Kerr metric (\ref{kere}) and 
$\mathcal{H}_{ij}^k$ are explicitly written as
\allowdisplaybreaks
\begin{align}
    \mathcal{H}_{tt}^0=& 2 r_h^2 \left(\frac{3 r^3}{r_h^3}+\frac{2 \left(2-\frac{r}{r_h}\right)}{\frac{r^2}{r_h^2}+1}-3 \left(\frac{r^2}{r_h^2}-1\right)^2 \cot ^{-1}\left(\frac{r}{r_h}\right)-\frac{7 r}{r_h}+3\right),\nonumber\\
    \mathcal{H}_{tt}^2=& \frac{ r_h^2 \left(\frac{r}{r_h}-1\right)^2}{ \left(\frac{r^2}{r_h^2}+1\right)} \left(-\frac{45 r^5}{r_h^5}-\frac{150 r^4}{r_h^4}-\frac{150 r^3}{r_h^3}-\frac{222 r^2}{r_h^2}\right.\nonumber\\
    &\left.+15 \left(\frac{3 r^6}{r_h^6}+\frac{10 r^5}{r_h^5}+\frac{11 r^4}{r_h^4}+\frac{18 r^3}{r_h^3}+\frac{9 r^2}{r_h^2}+\frac{8 r}{r_h}+1\right) \tan ^{-1}\left(\frac{r_h}{r}\right)-\frac{85 r}{r_h}-62\right),\nonumber\\
    \mathcal{H}_{t\phi}^0=&-8 r_h^3 \left(\frac{r}{r_h}-1\right)^2\!\! \left(-\frac{r}{r_h} \left(\frac{4 r}{r_h}+3\right)+ \left(\frac{r^2}{r_h^2} \left(\frac{4 r}{r_h}+3\right)+4\frac{r}{r_h}+1\right) \cot ^{-1}\!\left(\frac{r}{r_h}\right)\!-\!3\right),\nonumber\\
    \mathcal{H}_{t\phi}^2=&4 r_h^3 \left(\frac{r}{r_h}-1\right)^2 \left(-\frac{5 r}{r_h} \left(\frac{18 r^3}{r_h^3}+\frac{15 r^2}{r_h^2}+\frac{22 r}{r_h}+7\right)\right.\nonumber\\&\left.+5 \left(\frac{18 r^5}{r_h^5}+\frac{15 r^4}{r_h^4}+\frac{28 r^3}{r_h^3}+\frac{12 r^2}{r_h^2}+\frac{10 r}{r_h}+1\right) \cot ^{-1}\left(\frac{r}{r_h}\right)-22\right),
    \nonumber\\
    \mathcal{H}_{\phi\phi}^0=&\frac{1}{2} r_h^4 \left(\frac{r}{r_h}-1\right)^2 \left(\frac{r}{r_h} \left(-\frac{15 r^4}{r_h^4}+\frac{50 r^3}{r_h^3}+\frac{20 r^2}{r_h^2}+\frac{84 r}{r_h}+27\right)\right.\nonumber\\&\left.+5 \left(\frac{3 r^4}{r_h^4}-\frac{10 r^3}{r_h^3}-\frac{6 r^2}{r_h^2}-\frac{10 r}{r_h}-1\right) \left(\frac{r^2}{r_h^2}+1\right) \cot ^{-1}\left(\frac{r}{r_h}\right)+34\right),\nonumber\\
    \mathcal{H}_{\phi\phi}^2=&\frac{1}{7} r_h^4 \left(\frac{r}{r_h}-1\right)^2 \!\!\left(\frac{315 r^7}{r_h^7}-\frac{1470 r^6}{r_h^6}-\frac{735 r^5}{r_h^5}-\frac{3220 r^4}{r_h^4}-\frac{1407 r^3}{r_h^3}-\frac{2074 r^2}{r_h^2}\!-\!\frac{381 r}{r_h}\right.\nonumber\\
    &\left.\!-\!324\!-\!35 \left(\frac{r^2}{r_h^2}+1\right) \!\left(\frac{9 r^6}{r_h^6}-\frac{42 r^5}{r_h^5}-\frac{27 r^4}{r_h^4}-\frac{64 r^3}{r_h^3}-\frac{21 r^2}{r_h^2}-\frac{22 r}{r_h}-1\right) \cot ^{-1}\left(\frac{r}{r_h}\right)\right),
    \nonumber\\
    \mathcal{H}_{\theta\theta}^0=&-\frac{8}{15} r_h^4 (5 \frac{r}{r_h}-2 ) (\frac{r}{r_h}-1)^2,
    \nonumber\\
    \mathcal{H}_{\theta\theta}^2=&-\frac{2}{21}  r_h^4 \left(\frac{21 r^2}{r_h^2}-\frac{28 r}{r_h}-17\right) \left(\frac{r}{r_h}-1\right)^2,
    \nonumber\\
\mathcal{H}_{rr}^k=&\frac{\mathcal{H}_{\theta\theta}^k}{\Delta(r)}, \quad k=0,2 \ .
 \label{HH}   
\end{align}
\section{Other choices of multipole basis}
In our proof we have taken advantage of the fact that the Laplace equation for $\psi$ (\ref{psiab}) can be rewritten in Weyl-Shperical coordinates $(R,u,\phi)$ and still admit multipoles as solutions. On the other hand someone could easily be tempted in following a different approach by solving (\ref{psiab}) as it stands (in prolate coordinates). These set of solutions has in fact already been studied by ~\cite{Castejon-Amenedo:1990yso} . Although the resultant $\psi$ function describing a rotating spacetime in an Erez-Rozen gravitational field is quite friendly:
\begin{align}
    \psi = \sum_{n=0}^{\infty} (-1)^{n+1} q_n Q_n(x) P_n(y) \label{ERsol}
\end{align}
The emerged functions $a(x,y),b(x,y)$ and even $\gamma_s$ have the following forms:
\begin{align}
    a =& -\alpha \exp \bigg\{ 2\alpha \sum_{l=1}^{\infty} (-1)^l q_l D_- \bigg\} ,\nonumber\\
    b =& \alpha \exp \bigg\{ 2\alpha \sum_{l=1}^{\infty} (-1)^l q_l D_+ \bigg\} ,\nonumber\\
    \gamma_s =& \alpha^2\sum_{m,n=0}^{\infty} (-1)^{m+n} q_m q_n \int_{-1}^y \Gamma_y^{mn} \, dy,
\end{align}
where ,
\begin{align}
     D_{\mp} =& \frac{1}{2} (\pm 1)^l \ln \left( \frac{\left( x \mp y \right)^2}{x^2 - 1} \right)- (\pm 1)^l Q_1 + P_l Q_{l-1} - \sum_{k=1}^{l-1} (\pm 1)^k P_{l-k} \left( Q_{l-k+1} - Q_{l-k-1} \right),\nonumber\\ 
    \Gamma_y^{mn} =& (x^2 - 1) P_m' Q_m \left( 2x P_n Q_n' - y P_n' Q_n \right) \nonumber\\
    &+ \frac{(x^2 - 1)^2}{x^2 - y^2} \left[ P_m Q_m' (y P_n Q_n' - x P_n' Q_n) \right. 
    \left. + P_m' Q_m (y P_n' Q_n - x P_n Q_n') \right],
\end{align}
where the following formalism was used
\begin{align}
    P_n=P_n(x), \quad Q_n=Q_n(y), \quad P_n' = \frac{d P_n(y)}{dy} , \quad Q_n' = \frac{d Q_n(x)}{dx}
\end{align}
Therefore we realize that this formalism creates complicated and not enlightening expressions when someone has to deal with $n=2$ multipoles and beyond (as we did in our study). Because of that we preferred Weyl multipoles instead of the Erez-Rosen solution (\ref{ERsol}). 
\bibliographystyle{JHEP}
\bibliography{KERR}
\end{document}